\documentstyle[12pt]{article}    
\begin{document}
\begin{center}
   
        {\bf   Casimir Effect on the Radius Stabilization of the
Noncommutative Torus}

\vspace{1cm}

                      Wung-Hong Huang\\
                       Department of Physics\\
                       National Cheng Kung University\\
                       Tainan,70101,Taiwan\\

\end{center}
\vspace{2cm}

   We evaluate the one-loop correction to the spectrum of Kaluza-Klein
system for the $\phi^3$ model on $R^{1,d}\times (T_\theta^2)^L$, where
$1+d$ dimensions are the ordinary flat Minkowski spacetimes and the extra
dimensions are the L two-dimensional noncommutative tori with
noncommutativity $\theta$.  The correction to the Kaluza-Klein mass
spectrum is then used to compute the Casimir energy.    The results show
that when $L>2$ the Casimir energy due to the noncommutativity could give
repulsive force to stabilize the extra noncommutative tori in the cases of
$d = 4n - 2$, with $n$ a positive integral.    

\vspace{3cm}

\begin{flushleft}
   
E-mail:  whhwung@mail.ncku.edu.tw\\

\end{flushleft}

\newpage
\section  {Introduction}

The Casimir effect is the contribution of a non-trivial geometry on the
vacuum fluctuation [1-3].  The corresponding change in the vacuum electric
energy is found to attract the two perfectly conducting parallel plates in
the original investigation [1].   The property of the attractive Casimir
force is also found in  the gravitational system [4].   It is proposed to
render the extra spaces in the Kaluza-Klein unified theory [5] sufficiently
small and thus could not be observed.  It is generally believed that the
nonperturbative quantum gravity can stabilize the size of the extra spaces,
as the minimum length scale, Plank scale, is set in.   

   Another available scale is the noncommutativity $\theta_{ij}$ revealed
in the string theory [6-8].    It is proved to arise naturally in the
string/M theories [8].    Initially,  Connes, Douglas and Schwarz [6]  had
shown that the supersymmetric gauge theory on noncommutative torus is
naturally related to the compactification of Matrix theory [8].   More
recently,  it is known that the dynamics of a D-brane [9] in the presence
of  a B-field can, in certain limits, be described by the noncommutative
field theories [7].   

A distinct characteristic of the  noncommutative field theories, found by
Minwalla, Raamsdonk and Seiberg [10], is the mixing of ultraviolet (UV) and
infrared (IR) divergences reminiscent of the UV/IR connection of the string
theory.   In a recent paper [11] we found that the noncommutativity does
not affect one-loop effective potential of the scalar field theory.   
However, it can become dominant in the two-loop level and have an
inclination to induce the spontaneously symmetry breaking if it is not
broken in the tree level, and have an inclination to restore the symmetry
breaking if it has been broken in the tree level.  In the paper [12] Nam
has tried to use the noncommutativity as a  minimum scale to protect the
collapse of the extra spaces.  He use the one-loop Kaluza-Klein spectrum
derived by Gomis, Mehen and Wise [13] to compute the one loop Casimir
energy of an interacting scalar field in a compact noncommutative space of
$R^{1,d}\times T^2_\theta$, where $1+d$ dimensions are the ordinary flat
Minkowski space and the extra two dimensions are noncommutative torus with
noncommutativity $\theta$.   The correction is found to contribute an
attractive force and have a quantum instability.  He therefore turns to
investigate the case of vector field and find the repulsive force for
$d>5$.   We will mention the mistake in there and the above conclusion is
thus unreliable.

In this paper we extend the problem to the theory on $R^{1,d}\times
(T^2_\theta)^L$ while consider only the scalar field theory, i.e., $\phi^3$
theory with extra $L$ two-dimensional noncommutative tori.  In section II,
we first extend the works of Gomis, Mehen and Wise [13] to evaluate the
one-loop correction to the spectrum of Kaluza-Klein system for the $\phi^3$
model on $R^{1,d}\times (T_\theta^2)^L$.  The correction to the
Kaluza-Klein mass formula has the additional term which resembles that of
the winding states in the string theory [14], likes as the property found
in the $L=1$ system [13].   In section III, the obtained spectrum is used
to compute the Casimir energy.  Section IV is used to analyze the effect of
the correction to the Casimir energy on the radius stabilization.   We show
that when $L>2$, the Casimir energy in the cases of $d = 4n - 2$, with $n$
a positive integral, could give repulsive force to stabilize the extra
noncommutative tori.    This thus suggests a possible stabilization
mechanism for a scenario in Kaluza-Klein theory, where some of the extra
dimensions are noncommutative. 


\section {Kaluza-Klein Spectrum}

We consider the noncommutative scalar $\phi^3$ theory in $R^{1,d} \times
(T^2_\theta)^L$ spacetime described by the following action:

$$ S = \int d^{1+d} x ~ d^{2L}y \left( {1\over 2} (\partial \phi)^2
-{1\over 2} m^2 \phi^2 - { \lambda \over 3!} \phi \star \phi \star
\phi\right). \eqno{(2.1)}$$ 

\noindent
The $\star$ operator is the Moyal product generally defined by [10]

$$f(x) \star  g(x) = e^{+{i\over 2} \theta^{\mu\nu} {\partial\over \partial
y^\mu} 
{\partial\over \partial z^ \nu} } f(y) g(z) |_{y,z\rightarrow x}.  
\eqno{(2.2)} $$
\noindent
in which $\theta_{\mu\nu}$ is a real, antisymmetric matrix which represents
the noncommutativity of the spacetime, i.e., $ [x^\mu,x^\nu] = i \theta
^{\mu \nu}$. 
In Eq.(2.1) the coordinates $x^0,x^1,..., x^d$ represent the commutative
four dimensional Minkowski spacetime.  The $2L$ extra dimensions are taken
to be the $L$ noncommutative 2-tori $T^2_\theta$ whose noncommutative
coordinates are described by

$$ [y^1,y^2] =[y^3,y^4] = ... = [y^{2L-1},y^{2L}] = i \theta.  
\eqno{(2.3)}$$ 

\noindent
When $L=1$, this coordinate can be realized in string theory by wrapping a
five-brane on a two-torus $T^2$ with a constant $B$-field along the torus. 
The low energy effective four dimensional theory resulting from
compactification on a noncommutative space is local and Lorentz invariant,
hence it can be relevant phenomenologically [13]. 

   The momentum in the $1+d$ Minkowski spacetime, denoted as $p$, is a
continuous variable.   However, the momenta along the tori are quantized as
$\vec k \to \vec k /R$, where $\vec k = (k_1 ..., k_{2L})$ are the
integrals.  Therefore, using the Feynman rule [13], which  includes the
propagator

\unitlength 2mm
\begin {picture}(30,5)
\put(15,0){\vector(1,0){5}}
\put(20,0){\line(1,0){5}}
\put(18,1){$p, \vec n$}
\end {picture}
\hspace{5mm} $\frac {i (1- \delta_{\vec n ,0})}{p^2 - {\vec n}^2 - m^2}$

\noindent
and vertex

\begin {picture}(50,12)
\put(15,0){\line(1,1){5}}
\put(20,5){\line(1,-1){5}}
\put(20,5){\line(0,1){5}}
\put(17,0){$\vec k$}
\put(22,0){$\vec n$}
\put(21,8){$\vec m$}
\put(30,5){$ - i \lambda ~ cos( \frac {\theta}{2R^2} \vec n \wedge \vec k)
\delta_{\vec k +\vec n +\vec m,0}$}
\end {picture}

\noindent
in which $\vec{n} \wedge \vec{k} \equiv (n_1 k_2 - n_2 k_1) +(n_3 k_4 - n_4
k_3) ...+ (n_{2L-1} k_{2L} - n_{2L} k_{2L-1})$, the one loop contributions
to the two point functions is  

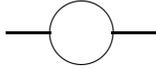
\begin {picture}(14,8)
\put(7,2){\circle{4}}
\put(2,2){\line (1,0 ){3}}
\put(9,2){\line (1,0 ){3}}
\end {picture}

$$ ={\lambda^2 \over 4} {1\over (2 \pi R)^{2L}} \sum_{\vec{k}} \int
{d^{1+d} l
\over (2 \pi)^{1+d}} {\cos^2(\theta ~ \vec{n} \wedge \vec{k}/(2 R^2) )(1-
\delta_{\vec{k},0}) (1-\delta_{\vec{n}+\vec{k},0})\over (l^2 -\vec{k}^2/R^2
- m^2) ((l+p)^2 - (\vec{n}+\vec{k})^2/R^2 - m^2)}  $$

$$={\lambda^2 \over 4} {1\over (2 \pi R)^{2L}} \sum_{\vec{k}} \int {d^{1+d}
l
\over (2 \pi)^{1+d}} {1-2\, \delta_{\vec{k},0} -
2\delta_{\vec{n}+\vec{k},0}+\cos( \theta ~ \vec{n} \wedge \vec{k}/R^2) 
\over (l^2 -\vec{k}^2/R^2 - m^2) ((l+p)^2 -(\vec{n} +\vec{k})^2/R^2 - m^2)
}.    \eqno{(2.4)}$$

\noindent
In which $l$ denotes the loop momenta along the noncompact directions,
while $\vec{k}/R$ is loop momenta along compact directions. Similarly, $p\,
(\vec{n}/R)$ is the external momenta along the noncompact(compact)
directions.   The derivation of Eq.(2.4) has used the half angle formula
for the cosine and the property that $\vec{n} \wedge \vec{n} = 0$, as
detailed by Gomis, Mehen and Wise [13].

The first term, second term and third term in Eq.(2.4) have divergences
which 
can be absorbed by the counterterms [13].  The corrected spectrum
calculated in Eqs.(2.8) and (2.9), coming from the last term, shows a
factor ${1 \over \theta^{2(L+{d-3\over 2 })}}$.    This means that  the
last term will become the leading contribution for small $\theta$.  
Therefore the first term, second term and third term are irrelevant to our
investigation and will be not discussed furthermore.   The last term
contains a oscillatory factor $\cos (\theta \,\vec{n} \wedge \vec{k}/R^2)$
which makes the non-planar correction term to be ultraviolet finite and
give the leading behavior for small $\theta$.  It will be evaluated in
below.

   We use the Feynman parameter $x$ to perform the integral over the
momentum $l$ and then express the result in terms of the Schwinger
parameter $\alpha$.   The one-loop self energy evaluated from Eq.(2.4) then
becomes 

$$\Sigma = - \frac {\lambda^2} {4 (4 \pi )^{L+\frac{1+d}{2}} } \int_0^1\ dx
\int_0^\infty\ d\alpha ~ \alpha^{1-L-\frac{1+d}{2}} exp \left[-\alpha
[m^2+x(1-x)  (- p^2+\frac{{\vec n} ^2} {R^2})] - \frac{\theta^2 {\vec n}
^2}
{4\alpha R^2}\right] \times$$
$$\left[\frac{1}{2} ~\Pi_{i=1}^{2L-1}~  \vartheta (x n_i + i \frac{\theta
n_{i+1}}{2 \alpha}) \vartheta (x n_{i+1} - i \frac{\theta n_{i}}{2 \alpha})
+ \frac{1}{2} ~\Pi_{i=1}^{2L-1}~  \vartheta (x n_i - i \frac{\theta
n_{i+1}}{2 \alpha}) \vartheta (x n_{i+1} + i \frac{\theta n_{i}}{2
\alpha})\right]. $$
                         $$ \eqno{(2.5)}$$
\noindent
To obtain the above result  we have  performed the sum over $\vec{k}$ by
using the definition of the Jacobi theta function 
$$ \vartheta(\nu, \tau) = \sum_{n=-\infty}^{\infty} \exp(\pi i n^2 \tau +2
i\pi n \nu),    \eqno{(2.6)}$$ 
and the property of modular transformation
$$\vartheta(\nu,\tau)=(-i \tau)^{-1/2}\exp( -\pi i \nu^2/\tau)
\vartheta(\nu/\tau, -1/\tau).    \eqno{(2.7)}$$

We see that the ultraviolate divergent contribution of the one-loop self
energy Eq.(2.5) comes from the $\alpha \to 0$ region [13].    Thus, to
obtain the leading contribution of the one-loop self energy, we can
approximate  $\vartheta =1$ in the Eq.(2.5) and the leading correction to
the spectrum of Kaluza-Klein system becomes

$$\Sigma = - {\lambda_{\theta}}^2 ~ \left (\frac {R^2}{{\vec n}
^2}\right)^{L+\frac{d-3}{2}}, \eqno{(2.8)}$$ 
in which 
$$ {\lambda^2_{\theta}}  =  \frac {\lambda^2 ~  \Gamma
({L+\frac{d-3}{2}})} {64 ~ \pi ^{L+ \frac{d+1} {2}} ~ \theta^{
2({L+\frac{d-3}{2}})}} .  \eqno{(2.9)}$$

\noindent
The result reduces to that in [13] when $L=1, d=3$.   

Note that the above result tells us that the correction term will depend on
the dimension $d$.   However, the investigation in [12] use the form of
$d=3$ to study the cases in other dimensions and thus draw the wrong
conclusion.   The same mistake also appears in [12] in investigating the
vector field system.  


\section {Casimir Energy}

Casimir energy is obtained by summing up the energy ~ $\omega$ ~ of all the
modes as follows [1]:

$$u = {1\over 2} \sum_{n_1,...,n_{2L}=1 }^{\infty}\int {d^d \vec{p} \over
(2\pi)^d} ~ \omega_{\vec n, \vec p}  \hspace{7cm}$$

$$ = {1\over 2} \sum_{n_1,...,n_{2L}=1 }^{\infty}\int {d^d \vec{p} \over
(2\pi)^d} \sqrt {{\vec p}^2 + {\vec{n}^2\over  R^2} -
{\lambda^2_{\theta}} ~ (\frac { R^2}{{\vec n}^2})^{L+\frac{d-3}{2}}}
\hspace{2.5cm}$$
$$= {1\over 2} \sum_{n_1,...,n_{2L}=1 }^{\infty} \int {d^d \vec {p} \over
(2\pi)^d }\int^\infty_0 {dt\over t} t^{-1/2}e^{-t \left({\vec p}^2 +
{\vec{n}^2\over  R^2} - {\lambda^2_{\theta}} ~ (\frac { R^2}{{\vec
n}^2})^{L+\frac{d-3}{2}}    \right)}  \hspace{1cm}$$

$$= {-1\over 4 \sqrt{\pi}}{1\over (4\pi   )^{d/2}}  \sum_{n_1,...,n_{2L}=1
}^{\infty} \int^\infty_0 {dt\over t} t^{-(d+1)/2}e^{-t
\left({\vec{n}^2\over  R^2} - {\lambda^2_{\theta}} ~ (\frac { R^2}{{\vec
n}^2})^{L+\frac{d-3}{2}}    \right)}  \hspace{1cm}  $$
$$= {-1\over 4 \sqrt{\pi}}{1\over (4\pi   )^{d/2}}  \Gamma(-
\frac{d+1}{2})\sum_{n_1,...,n_{2L}=1 }^{\infty}  \left({\vec{n}^2\over 
R^2} - {\lambda^2_{\theta}} ~ (\frac { R^2}{{\vec
n}^2})^{L+\frac{d-3}{2}}\right)^{\frac{d+1}{2}}, \eqno{(3.1)}   $$

\noindent
in which, for simplicity, we consider the massless case.   To obtain the
above result we first use the Schwinger's proper time $t$ to handle the
square root, then integrate the transverse momentum $\vec p$  by doing a
Gaussian integral, and finally  integrate the Schwinger's proper time by
using the integral representation of Gamma function.

In the perturbative regime, ${\lambda^2_{\theta}} (\frac { R^2}{{\vec
n}^2})^{L+\frac{d-3}{2}} << 1$, we can approximate Eq.(3.1) by

$$u= {-1\over 4 \sqrt{\pi}}{1\over (4\pi   )^{d/2}}  \Gamma(-
\frac{d+1}{2}) \sum_{n_1,...,n_{2L}=1 }^{\infty}  \left({\vec{n}^2\over 
R^2}\right)^{\frac{d+1}{2}}\left [ 1 - \frac{d+1}{2} {\lambda^2_{\theta}}
~ (\frac { R^2}{{\vec n}^2})^{L+\frac{d-3}{2}} + ... \right] \hspace{1cm}$$

$$= {-1\over 4 \sqrt{\pi}}{1\over (4\pi   )^{d/2}}\Gamma(-
\frac{d+1}{2})\left [ v_{2L}(- \frac{d+1}{2}) \frac{1}{R^{d+1}} -  
{\lambda^2_{\theta}}\frac{d+1}{2}v_{2L} (L-1) R^{2(L-1)} \right],
\eqno{(3.2)}$$
in which 
$$v_N(s) \equiv \sum_{n_1,...,n_{N}=1 }^{\infty}\left[ {1\over n_1^2
+...+n_N^2} \right]^{-s} . \eqno{(3.3)}$$

From Eq.(3.2) and reflection formula [2]
$$\pi^{-s} \Gamma (s) v_N(s) = \pi^{s-{N\over 2}} \Gamma ({N\over 2}-s)
v_N({N\over 2}-s),        \eqno{(3.4)}$$
\noindent
we finally find that the Casimir energy can be expressed as 

$$u = {-1\over 4 \sqrt{\pi}}{1\over (4\pi   )^{d/2}} [{1\over \pi^{L+d+1}}
\Gamma(L+ \frac{d+1}{2}) ~ v_{2L}(L+\frac{d+1}{2})\frac{1}{R^{d+1}} - $$
$$ {\lambda^2_{\theta}} \frac{d+1}{2} \Gamma(- \frac{d+1}{2}) ~
v_{2L}(L-1) R^{2(L-1)} ]. \eqno{(3.5)}$$
\\
The first term is negative and contribute an attractive force to
unstabilize the radius.    When the second term, which depends on the
noncommutativity $\theta$, is positive then the correction due to the space
noncommutativity may stabilize the radius.  The compactification radius, if
it exists,  will be at

$$ R=\left[\frac{-\Gamma(L+ \frac{d+1}{2}) v_{2L}(L+\frac{d+1}{2})
}{{\lambda^2_{\theta}} \pi^{L+d+1}(L-1) \Gamma(- \frac{d+1}{2})v_{2L} 
}\right]^{1\over{2L+d-1}} .\eqno{(3.6)}$$
\\
\noindent
In the next section we will use the above results to analyze the
stabilization of the noncommutative extra tori.  


\section {Results and Conclusions}
$(a) ~ L=1$:

Let us first analyze  the case of  $L=1$.   From Eq.(3.5) we see that when
$L=1$, i.e., extra noncommutative space is a single 2-torus,  and $d$ is
odd, then the contribution from the correction to the Casimir energy is
finite after using the reflection formula Eq.(3.4).    The correction term
is independent of the radius $R$ of the torus.   Thus, up to the order of
perturbation there is no stabilization and we have to consider the next
order in correction to the Casimir energy.  This is contrast to that in
[12].   The investigation in [12] use the spectrum of $d=3$ to study the
Casimir effect in other dimensions and thus draws a wrong conclusion that
when $d>3$ the correction will have attractive force to unstabilize the
size of radius $R$. 

   Note that when $d$ is a odd number, then the divergent term
$\Gamma (- \frac{d+1}{2}) $ in Eq.(3.5)  will lead the Casimir energy to be
 infinite.  This may mean that the approximation used to derive Eq.(3.2)
from Eq.(3.1) is broken down.    This seems something strange and the
problem is remained to be solved. 

\noindent
$(c) ~ L=2$:

In this case $v_{4}(1)$ is divergent. Thus the approximation used to derive
Eq.(3.2) from Eq.(3.1) is broken down.   The problem is remained to be
solved.

\noindent
$(c) ~ L>2$:

Because there is no stabilization for a single or two 2-torus up to the
order of
perturbation, let us turn to the space with more 2-torus. From Eq.(3.5)
we see that when $d = 4 n$, with $n$ an integral, then $\Gamma (-
\frac{d+1}{2}) > 0$ and the correction energy is negative.  Thus the
correction term will contribute a attractive force and system does not have
a stable radius. However, when $d = 4 n - 2$, with $n$ a positive integral,
then $\Gamma (- \frac{d+1}{2}) < 0$ and the correction energy is positive. 
 Thus the system will have a stable radius.   The compactification radius
is expressed in the formula Eq.(3.6).

   Note that when $L>2$ and $d$ is an odd number, then the divergent term
$\Gamma (- \frac{d+1}{2}) $ in Eq.(3.5) will lead the Casimir energy to be
 infinite.   Just likes that in the case of $L =1$ with odd $d$, this may
mean that the approximation used to derive Eq.(3.2) from Eq.(3.1) is broken
down.  The problem is remained to be solved.

    Finally, let us make some remarks:

(1)  In this paper we have seen that when $L>2$ the Casimir energy of
$\phi$ in the cases of  $d =4n -2$, with $n$ a positive integral, could
give repulsive force to stabilize the extra noncommutative tori.  This thus
suggests a possible stabilization mechanism for a scenario in Kaluza-Klein
theory, where some of the extra dimensions are noncommutative.    

(2)  The $\phi^3$ interaction considered in this paper is like the
interaction in the string field theory [15].    Therefore it has some
motivations from the string theory despite of the fact that the $\phi^3$
theory itself will become divergent if the spacetime is larger then six.  
However, when the spacetime is larger then six we can assume that some
fields, coming from the other modes of the string, will contribute to the
loop diagram and  make the total system renormalizable.    We have in the
section III, therefore, calculated the Casimir effect of the $\phi$ field
while neglect others. 

(3)   The Casimir effect is null in the supersymmetry system, as the
contribution of boson field is just canceled by that of the fermion field .
  However, some mechanisms are proposed to break the supersymmetry to
describe the physical phenomena.   Thus the remaining Casimir effect may be
used  to render the compact space stable.  An interesting mechanism to
break the supersymmetry is the temperature effect, which is the scenario in
the early epoch of the universe.   Therefore it is useful to investigate
the finite-temperature Casimir effect of $\phi^3$ model on $R^{1,d}\times
(T_\theta^2)^L$.   It will be presented in elsewhere [16].  

\newpage

\begin{enumerate}
\item H.B.G. Casimir,{\it ``On the Attraction Between Two Perfectly
Conducting Plates''}, Proc. K. Ned. Akad. Wet, {\bf 51} (1948) 793; C.
Itzykson and J. B. Zuber, {\it Quantum Field Theory}, New York,
McGraw-Hill, 1980.

\item J. Ambjorn and S. Wolfram, Ann. Phys. {\bf 147} (1983) 1.
\item G. Pluien, B. M\"uller, and W. Greiner, Phys. Rep. {\bf 134} (1986)
87;\\
 V.M. Mostepanenko and N.N. Trunov, {\it ``The Casimir Effect and its
Applications''}, Oxford Univ. Press, 1997.

\item T. Appelquist and A. Chodos, Phys. Rev. Lett.{\bf 50} (1983) 141;
Phys. Rev. {\bf D28} (1983) 772. 

\item   Th. Kaluza, Situngsber. d. K. Preuss. Akad. d. Wissen.
z. Berlin, Phys.-Math. Klasse (1921) 966; O. Klein, Z. F. Physik 37 (1926)
895;\\
T. Appelquist, A. Chodos, and P.G.O. Freund,``{\it Modern Kaluza-Klein
Theories}'', Addison-Wesley, Menlo Park,  1987.

\item  A. Connes, M. R. Douglas and A. Schwarz,
  ``Noncommutative Geometry and Matrix Theory: Compactification on
  Tori'', hep-th/9711162, JHEP 9802:003 (1998); \\
 B.~Morariu and B.~Zumino, ``Super Yang-Mills on the
  Noncommutative Torus'', hep-th/9807198; \\
 C.~Hofman and E.~Verlinde, ``U-duality of Born-Infeld on the
Noncommutative Two-Torus'',  hep-th/9810116, JHEP {\bf 9812}, 010 (1998).

\item  N.~Seiberg and E.~Witten,
``String Theory and Noncommutative Geometry'',  hep-th/9908142, 
JHEP {\bf 9909}, 032 (1999).

\item N. A. Obers and B. Pioline, Phys. Rep. {\bf 318} (1999) 113,
hep-th/9809039. 

\item  J. Polchinski, {\it String Theory}, Cambridge University Press,
1998.

\item S.~Minwalla, M.~Van Raamsdonk and N.~Seiberg,
``Noncommutative Perturbative Dynamics'',
hep-th/9912072, JHEP {\bf 0002}, 020 (2000);\\
     C.~P.~Martin and D.~Sanchez-Ruiz,``The One-loop UV Divergent Structure
of U(1) Yang-Mills Theory on Noncommutative $R^4$'', hep-th/9903077, Phys.\
Rev.\ Lett.\  {\bf 83}, 476 (1999);\\
 I.~Chepelev and R.~Roiban,``Renormalization of Quantum field Theories on
Noncommutative $R^d$. I: Scalars'', hep-th/9911098;\\ 
B. A. Campbell and K. Kaminsky, ``Noncommutative Field Theory and
Spontaneous Symmetry Breaking'', Nucl. Phys. B 581 (2000) 240,
hep-th/0003137;\\
A. Matusis, L Ausskind and  N. Toumbas, ``The IR/UV Connection in 
Noncommutative Gauge Theories'', hep-th/0002075 ;

\item W. H. Huang,  ``Two-Loop Effective Potential in Noncommutative scalar
field theory'',  hep-th/0009067.

\item S. Nam,  ``Casimir Force in Compact Noncommutative Extra Dimensions
and Radius Stabilization'',  hep-th/0008083.

\item J. Gomis, T. Mehen and M.B. Wise, ``Quantum Field Theories with
Compact Noncommutative Extra Dimensions'', JHEP {\bf 0008}, 029 (2000).
 hep-th/0006160. 

\item W. Fischler, E. Gorbatov, A. Kashani-Poor, S. Paban, P. Pouliot and
J. Gomis, ``Evidence for Winding States in Noncommutative Quantum Field
Theory'',  hep-th/0002067. 

\item M. B. Green, J. H. Schwarz and E. Witten, {\it Superstring Theory},
Cambridge University Press, 1986.

\item W. H. Huang,  ``Finite-temperature Casimir Effect on the Radius
Stabilization of the Noncommutative Torus'',  in preparation.

\end{enumerate}
\end{document}